\documentclass[a4paper,12pt]{article}
\usepackage{latexsym,amssymb,amsfonts,amsmath,amsthm}
\usepackage[dvipdfmx]{color}
\usepackage[dvipdfmx]{graphicx}
\usepackage{color}

\newtheorem{theorem}{Theorem}
\newtheorem{lemma}{Lemma}
\newtheorem{corollary}{Corollary}
\newtheorem{proposition}{Proposition}
\newtheorem{remark}{Remark}
\newtheorem{definition}{Definition}
\newtheorem{conjecture}{Conjecture}

\newcommand{\bs}[1]{\boldsymbol{#1}}
\newcommand{\im}{\bs{\rm i}}
\newcommand{\spann}{{\rm span}}

\newcommand{\e}{{\rm e}}

\title{{\Large {\bf How does Grover walk recognize the shape of crystal lattice?  
}
}}
\author{ 
{\small 
Chul Ki Ko,$^{1}$ 
\footnote{kochulki@yonsei.ac.kr 
}\quad 
Norio Konno,$^{2}$ 
\footnote{konno@ynu.ac.jp 
}\quad
Etsuo Segawa,$^{3}$ 
\footnote{
e-segawa@m.tohokku.ac.jp
}\quad
Hyun Jae Yoo $^{4}$ 
\footnote{yoohj@hknu.ac.kr 
}
}\\ 
{\scriptsize $^{1}$ 
University College, Yonsei University,
}\\
{\scriptsize 
85 Songdogwahak-ro, Yeonsu-gu, Incheon 21983, Korea.
} \\
{\scriptsize $^2$ 
Department of Applied Mathematics, Faculty of Engineering, Yokohama National University 
}\\
{\scriptsize 
Hodogaya, Yokohama 240-8501, Japan 
} \\
{\scriptsize $^3$ 
Graduate School of Information Sciences, Tohoku University, 
}\\
{\scriptsize 
Aoba, Sendai 980-8579, Japan
} \\
{\scriptsize $^4$ 
Department of Applied Mathematics, Hankyong National University,
}\\
{\scriptsize
327 Jungangro, Anseong-si, Gyeonggi-do 17579, Korea
}\\
} 
\date{\empty }
\begin{document}
\maketitle

\par\noindent
\begin{small}
\par\noindent
{\bf Abstract}. 
We consider the support of the limit distribution of the Grover walk on crystal lattices with the linear scaling. 
The orbit of the Grover walk is denoted by the parametric plot of the pseudo-velocity of the Grover walk in the wave space. 
The region of the orbit is the support of the limit distribution. 
In this paper, we compute the regions of the orbits for the triangular, hexagonal and kagome lattices. 
We show every outer frame of the support is described by an ellipse. 
The shape of the ellipse depends only on the realization of the fundamental lattice of the crystal lattice in $\mathbb{R}^2$. 


\end{small}

\setcounter{equation}{0}

\section{Introduction}
The Grover walk is one of the intensively studied mathematical models of quantum walks. 
This considerable reasons are as follows: 
(i) the Grover walk is a useful tool in the quantum computing accomplishing so called quantum speed up (see~\cite{Por} and its references therein); 
(ii) there is an underlying random walk which describes a part of the spectrum of the Grover walk~\cite{HKSS,Sze}; 
(iii) some stochastic behavior of the underlying random walk and also geometric aspect of the graph 
appear in a different forms as the limiting behavior of the induced Grover walk~\cite{HS,S};
(iv) there is a connection to some graph theoretical and combinatorial aspects 
inducing inverse problems to classify graphs by some Grover walk's behaviors~\cite{GG,HKSS2,Y}; 
(v) the Grover walk naturally appears as the potential-free quantum graph~\cite{GS}, 
which is a system of the Schr{\"o}dinger equation on the metric graph with the 
boundary conditions at the vertices preserving the self-adjointness of the Hamiltonian~\cite{T}. 

In this paper, we consider the Grover walk in the context of (iii) restricting the graphs to three typical crystal lattices; 
triangular, hexagonal and kagome lattices. 
In particular, we study the orbit of the Grover walker linearly scaled by the large time step $n$. 
The Grover walk on these graphs exhibits both localization and linear spreading~\cite{HKSS}. 
It is known that the orbit is described by the parametric plot of the group velocity in the wave space, 
and the density corresponding to how frequently the orbit of the Grover walker runs through each small mesh on $\mathbb{R}^2$ 
is expressed by the effective mass~\cite{WKKK}. 
We show that the orbit of the Grover walker is included in an ellipse with a rotation (Theorem~\ref{main}). 
The shape of the region depends only on the realization of the embedding of the fundamental lattice of the crystal lattice in $\mathbb{R}^2$. 
For the underlying random walk on the crystal lattices, 
geometric quantities and the realization of the embedding in $\mathbb{R}^d$ are reflected in the return probability~\cite{KSS,HS}. 
On the other hand, homological structure is reflected as the localization of the induced Grover walk~\cite{HKSS,HS}. 
However it has been still an open problem to find geometric properties of the graphs from the behavior of linear spreading of the Grover walk. 
Although this paper treats the orbits of only special crystal lattices linearly scaled by large time $n$, 
we expect that 
this is a first step to address to this open problem of the Grover walk and provide an interest of this problem. 

This paper is organized as follows. 
In section 2, the definition of the Grover walk on the connected graph is explained. 
Section 3 is devoted to a construction of the crystal lattice from the finite graph and an embedding of the crystal lattice in $\mathbb{R}^2$. 
In section 4, we prepare the setting of the Grover walk on the crystal lattice and take a short review on spectral mapping theorem and 
the limit theorem of this walk. We define the orbit of the Grover walk in this section. 
In section 5, we give our main theorem for the orbits of the Grover walk on the three crystal lattices and its proof. 
Finally, we discuss our conjecture of the orbits and verify it by numerical simulations.
\section{Definition of Grover walk on graph}
Let $G=(V(G),E(G))$ be a graph. We define $A(G)$ as the symmetric arcs induced by $E(G)$.  
The inverse arc of $e\in A(G)$ is denoted by $\bar{e}\in A(G)$. 
We denote $o(e),t(e)\in V(G)$ as the origin and terminal vertices of $e$, respectively. 
If $G$ is a simple graph, then the arc $e$ with $o(e)=u$ and $t(e)=v$ is denoted by $(u,v)$. 
The cycle $C$ is the sequence of arcs $(e_0,\dots,e_{r-1})$ so that $t(e_j)=o(e_{j+1})$ for every $j\in \mathbb{Z}/\mathbb{Z}_r$. 
The Grover walk on graph $G$ is defined as follows. 
\begin{definition}
\noindent
\begin{enumerate}
\item Total Hilbert space: $\mathcal{H}=\ell^2(A(G))=\{ \psi: A(G)\to \mathbb{C} \;: ||\psi||^2<\infty \}$. Here the inner product is the standard inner product, 
that is, $\langle \psi,\phi \rangle=\sum_{e\in A(G)} \overline{\psi(e)}\phi(e)$. 
\item Time evolution: $U: \mathcal{H}\to\mathcal{H}$ (unitary) such that
	\[  (U\psi)(e)=\sum_{f:o(e)=t(f)} \left( \frac{2}{\mathrm{deg}(o(e))}-\delta_{e,\bar{f}} \right)\psi(f)  \]
\item Finding probability at time $n$ is defined by $\mu_n^{(\psi_0)}: V(G)\to [0,1]$ with the initial state $\psi_0\in \ell^2(A(G))$ ($||\psi_0||=1$) 
such that
	\[ \mu_n^{(\psi_0)}(u)=\sum_{e:t(e)=u}| (U^n\psi_0)(e) |^2. \]
\end{enumerate}
\end{definition}

\section{Crystal lattice and its realization on $\mathbb{R}^d$}
Let $G_0=(V_0,E_0)$ be a finite graph which may have multi edges and self loops. 
We use the notation $A_0:=A(G_0)$ for the set of symmetric arcs induced by $E_0$. 
The homology group of $G_0$ with integer coefficients is denote by $H_1(G_0,\mathbb{Z})$. 
The abstract period lattice $L$ induced by a subgroup $H\subset H_1(G_0,\mathbb{Z})$ is denoted by $H_1(G_0,\mathbb{Z})/H$~\cite{Sunada}. 

Let the set of basis of $H_1(G_0,\mathbb{Z})$ be $\{C_1,C_2,\dots,C_{b_1}\}$ corresponding to fundamental cycles of $G_0$, 
where $b_1$ is the first Betti number of $G_0$. 
The spanning tree induced by $\{C_1,C_2,\dots,C_{b_1}\}$ is denoted by $\mathbb{T}_0$. 
We can take a one-to-one correspondence between $\{C_1,C_2,\dots,C_{b_1}\}$ and $A(\mathbb{T}_0)^{c}$; 
we describe $C(e)\in \{C_1,C_2,\dots,C_{b_1}\}$ as the fundamental cycle corresponding to $e\in A(\mathbb{T}_0)^{c}$ 
so that $C(e)$ is the cycle generated by adding $e$ to $\mathbb{T}$. 
Set $\phi: A_0 \to \mathbb{R}^d$ so that 
\begin{enumerate} 
\item for every $e\in A_0$, 
	\[ \phi(\bar{e})=-\phi(e); \]
\item 
	\[ \mathrm{rank}[\phi(C_1),\dots,\phi(C_{b_1})] =d. \]
Here for a fundamental cycle $C_j=(e_0,\dots,e_{r-1})$ ($j\in \{1,\dots,b_1\}$), we define 
	\[ \phi(C_j)=\phi(e_0)+\cdots+\phi(e_{r-1}). \]
\end{enumerate}
We also set $\phi_0: V_0\to \mathbb{R}^d$ so that 
	\[ \phi(e)=\phi_0(t(e))-\phi_0(o(e)) \]
for every $e\in A(\mathbb{T}_0)$. 
Thus the relative coordinate of each vertex of $G_0$ is determined by $\phi_0$. 
Remark that for every $e\in A(\mathbb{T}_0)^c$ corresponding to the fundamental cycle $C=(e_1,\dots,e_r,e)$, 
	\[ \phi(e)=\phi(C)+\{\phi_0(o(e_1))-\phi_0(t(e_r))\}=\phi(C)+\{\phi_0(t(e))-\phi_0(o(e))\}. \]

The covering graph $G=(V,A)$ of $G_0$ by the abstract period lattice $L$ is expressed as follows, where $A$ is the set of the symmetric arcs: 
	\begin{align*}
        V &= \phi(L)+\phi_0(V_0) \cong \phi(L)\times \phi_0(V_0); \\
        A= & \left(\bigcup_{x\in L} \left\{ \left((x,o(e)),(x,t(e))\right) \;|\; e\in A(\mathbb{T}_0) \right\}\right)  \\
        	& \quad \cup \left(\bigcup_{x\in L} \left\{ \left((x,o(e)),(x+\phi(C(e)),t(e))\right) \;|\; e\in A^{c}(\mathbb{T}_0) \right\}\right).
        \end{align*}
        
\section{Grover walk on the crystal lattice}
Let $\hat{\theta}: A_0\to \mathbb{R}^d$ be 
	 \[ \hat{\theta}(e)=\begin{cases} \phi(C(e)) & \text{: $e\in A^{c}(\mathbb{T}_0)$} \\ 0 & \text{: $e\in A(\mathbb{T}_0)$.}\end{cases} \]
It holds $\hat{\theta}(\bar{e})=-\hat{\theta}(e)$. 
We choose $e_{i_1},\dots,e_{i_d}$ from $A^c(\mathbb{T})$ so that 
$\hat{\theta}_1:=\hat{\theta}(e_{i_1}),\dots,\hat{\theta}_d:=\hat{\theta}(e_{i_d})$ span $\mathbb{R}^d$. 
We put the $d\times d$ matrix by
	\[ \Theta:={}^t([\hat{\theta}_1,\dots,\hat{\theta}_{d}]^{-1}). \]

\noindent{\bf Vertex based operator: twisted isotropic random walk}\\
Set the Hilbert space generated by $V=V(G)$ as $\ell^2(V)$. 
Recall that each vertex is represented by some $x\in L$ and $u\in V_0$ with $x+\phi_0(u)$. 
We shortly express $f(x+\phi_0(u))$ by $f(x;u)$. 
We define the Fourier transform with $k\in [0,2\pi)^d:=\mathcal{T}^d$ by 
	\[ \hat{f}(k;u) = \sum_{x\in L} f(x;u) \e^{\im \langle \Theta k, x \rangle}. \]
The inverse Fourier transform is expressed by
	\begin{align*} 
        f(x;u) &= \int_{\mathcal{T}^d} \hat{f}(k;u)\; \e^{-i\langle \Theta k,x \rangle} \frac{dk}{(2\pi)^d}, \\
        	&= \frac{1}{|\Theta|}\int_{\Theta(\mathcal{T}^d)} \hat{f}(\Theta^{-1}k;u)\; \e^{-\im \langle k, x\rangle} \frac{dk}{(2\pi)^d},
        \end{align*}
where $|\Theta|=\det (\Theta)$. 
The random walk operator on $G$ is described by
	\[ (Pf)(v)=\sum_{e\in A: t(e)=v}p(e)f(o(e)), \]
for every $v\in V$, where $p(e)=1/\deg (o(e))$.
Putting $\hat{P}_k: \ell^2(V_0)\to\ell^2(V_0)$ by 
	\[ (\hat{P}_kf_0)(u)= \sum_{e\in A_0:t(e)=u} p(e) \;e^{\im \langle \Theta k, \hat{\theta}(e)\rangle} f_0(o(e)), \]
we have 
	\begin{align*} 
        (P^nf)(x;u)  
        	&= \int_{\mathcal{T}^d} 
                \hat{P}_k^n \hat{f}(k;u)
                        \;\e^{-\im \langle \Theta k,x \rangle} \frac{dk}{(2\pi)^d}, \\
                &= \frac{1}{|\Theta|}\int_{\Theta(\mathcal{T}^d)} 
                \hat{P}_{\Theta^{-1}k}^n \hat{f}(\Theta^{-1}k;u)
                        \;\e^{-\im \langle k,x \rangle} \frac{dk}{(2\pi)^d}.        
        \end{align*}
Here $\hat{P}_k$ acts on $\hat{f}(k;\cdot)$  before taking the integration. 
We call $\hat{P}_k$ a twisted random walk on the quotient graph $G_0$. 
\\

\noindent{\bf Arc based operator: twisted Grover walk}\\
Set the Hilbert space generated by $A=A(G)$ as $\ell^2(A)$. 
Each arc is represented by some $x\in L$ and $e\in A_0$ with $(x;e)$. 
We define the Fourier transform with $k\in \mathcal{T}^d$ by 
	\[ \hat{\psi}(k;e) = \sum_{x\in L} \psi(x;e)e^{\im \langle \Theta k, x \rangle}. \]
The inverse Fourier transform is expressed by
	\begin{align*} 
        \psi(x;e) &= \int_{\mathcal{T}^d} \hat{\psi}(k;e)\; \e^{-i\langle \Theta k,x \rangle} \frac{dk}{(2\pi)^d}, \\
        	&= \frac{1}{|\Theta|}\int_{\Theta(\mathcal{T}^d)} \hat{\psi}(\Theta^{-1}k,u)\; \e^{-\im \langle k, x\rangle} \frac{dk}{(2\pi)^d},
        \end{align*}
where $|\Theta|=\det (\Theta)$. 
The Grover walk operator on $G$ is described by
	\[ (U\psi)(e)=\sum_{f\in A: o(e)=t(f)} \left(\frac{2}{\deg (o(e))}-\delta_{\bar{e},f}\right) \psi(f) \]
for every $e\in A$.
Putting $\hat{U}_k: \ell^2(A_0)\to\ell^2(A_0)$ by 
	\[ (\hat{U}_k\psi_0)(e) 
        	= \sum_{f\in A_0:t(f)=o(e)} \left(\frac{2}{\deg (o(e))}-\delta_{\bar{e},f}\right)
                	\;\e^{\im \langle \Theta k, \hat{\theta}(f)\rangle} \psi_0(f), \]
we have 
	\begin{align*} 
        (U^n\psi)(x;e)  
        	&= \int_{\mathcal{T}^d} 
                \hat{U}_k^n \hat{\psi}(k;e)
                        \;\e^{-\im \langle \Theta k,x \rangle} \frac{dk}{(2\pi)^d}, \\
                &= \frac{1}{|\Theta|}\int_{\Theta(\mathcal{T}^d)} 
                \hat{U}_{\Theta^{-1}k}^n \hat{\psi}(\Theta^{-1}k;e)
                        \;\e^{-\im \langle k,x \rangle} \frac{dk}{(2\pi)^d}.        
        \end{align*}
Here $\hat{U}_k$ acts on $\hat{\psi}(k;\cdot)$ before taking the integration. 
The unitary operator $\hat{U}_k$ is called a twisted Grover walk operator. 
\\

\noindent{\bf Via limit distribution} \\
A useful method to get the spectrum of $\hat{U}_k$ is obtained by \cite{HKSS}. 
\begin{theorem}\label{SpecMap}\cite{HKSS}(Spectral mapping theorem) Let $J(z)=(z+z^{-1})/2$. Then we have 
\[ \sigma(\hat{U}_k)\supseteq J^{-1}(\sigma(\hat{P}_k)); \; \sigma(\hat{U}_k)\setminus J^{-1}(\sigma(\hat{P}_k))\subseteq\{\pm 1\}. \]
Let $\mathcal{C}\subset \ell^2(A_0)$ be the eigenspace which is orthogonal to the eigenspace of $J^{-1}(\sigma(\hat{P}_k))$. Then
\begin{align*} 
\mathrm{dim} \ker(1-\hat{U}_k|_\mathcal{C}) &= b_1; \\
\mathrm{dim} \ker(1+\hat{U}_k|_\mathcal{C}) &= 
	\begin{cases} b_1 & \text{: $G_0$ is bipartite,} \\ b_1-1 & \text{: $G_0$ is non-bipartite,} \end{cases}
\end{align*}
where $b_1$ is the first Betti number of $G_0$.
\end{theorem}
\noindent \\
Let $\mu_n^{(\psi_0)}: A\to [0,1]$ be the probability measure at the $n$-iteration of $U$ with the initial state $\psi_0\in \ell^2(A)$ such that
	\[ \mu_n^{(\psi_0)}(x;e_0)=|(U^n\psi_0)(x;e_0)|^2. \]
Putting $e=(x;e_0)$, we take the Fourier transform of $\mu_n^{(\psi_0)}(\cdot \; ;e_0)$ for $\xi\in \mathbb{R}^d$ by 
	\[ \chi_n(\xi;e_0):= \sum_{x\in L} \mu_n^{(\psi_0)}(x;e_0)\;\e^{\im \langle \Theta \xi, x\rangle}.  \]
We have the following useful formula which connects the Fourier transform of the amplitude $\hat{\psi}$ 
and the Fourier transform of the probability $\chi_n$. 
\begin{proposition}
Let $\hat{\psi}_n=\hat{U}_k^n\hat{\psi}_0$. Then we have 
	\[ \chi_n(\xi;e_0)=\int_{\mathcal{T}^d} \overline{\hat{\psi}_n(k;e_0)}\hat{\psi}_n(k+\xi;e_0) \frac{dk}{(2\pi)^d}. \]
\end{proposition}
We join $\{\chi_n(\xi;e_0)\}_{e_0\in A_0}$ by 
	\[ \chi_n(\xi):=\sum_{e_o\in A_0}\chi_n(\xi;e_0). \]
This is rewritten by 
	\[ \chi_n(\xi)=\int_{\mathcal{T}^d} \mathrm{Tr}[ \hat{U}_{k+\xi}^n \;\rho_{k,\xi}^{(0)} \;\hat{U}_{k}^{-n}]  \frac{dk}{(2\pi)^d}, \]
where the matrix representation of $\rho_{k,\xi}^{(0)}$ is 
$(\rho_{k,\xi}^{(0)})_{e,f}=\hat{\psi}_0(k+\xi;e)\overline{\hat{\psi}_0(k;f)}$. 
As the initial state, we take the mixed state, i.e.,  $\rho_{k,\xi}^{(0)}=I_{|A_0|}/|A_0|$.
We put the characteristic function with this initial state by $\chi_n^{(o)}$. 
\begin{theorem}(\cite{HKSS})
Let the eigenvalues of $\hat{P}_k$ be $\{\cos\gamma_j(k)\}_{j=1}^{|V_0|}$ and 
we assume $\gamma_j\in C^{\infty}$.
	\begin{equation}\label{chara} 
        \lim_{n\to\infty}\chi_n^{(o)}(\xi/n)
        	=\frac{1}{|E_0|}\int_{\mathcal{T}^d} \; \sum_{j=1}^{|V_0|}\; \cos(\langle \xi, \nabla\gamma_j(k)\rangle) \frac{dk}{(2\pi)^d}
                +\frac{2b_1-\bs{1}_B}{2|E_0|}.  
        \end{equation}	
Here $\bs{1}_B=1$ if $G_0$ is non-bipartite, $\bs{1}_B=0$ if $G_0$ is bipartite. 
\end{theorem}
The existence of the second term of (\ref{chara}), which is independent of $\xi$, means that localization exhibits in this quantum walk. 
In this paper, we focus on the first term corresponding to the linear spreading of this quantum walk. 
If $\mathrm{H}_{\gamma_j}=\mathrm{det}[(\partial^2 \gamma_j/\partial k_\ell\partial k_m)_{\ell,m=1}^{d}]\neq 0$ for almost every $k\in \mathcal{T}^d$, then 
by replacing the variable $\nabla \gamma_j(k)$ into $x\in \mathbb{R}^d$ $(j\in \{1,\dots,|V_0|\})$, 
the first term of RHS for such a $j$ is expressed by 
	\[ \int_{x\in \mathbb{R}^d}  \e^{\im \langle \xi, x\rangle } \rho_j(x) dx \]
Here $\gamma_j(k)$ is decomposed into $\gamma_{j,1}(k)+\gamma_{j,2}(k)+\cdots+\gamma_{j,\kappa}(k)$ so that for each $\ell$, 
$\nabla \gamma_{j,\ell}(k)$ is in one to one correspondence with $x$ 
and  $\rho_j$ is given by $\rho_{j}= \sum_{\ell} (|\mathrm{H}_{\gamma_j}|^{-1})|_{\nabla \gamma_{j,\ell}(k)=x}$. 
\section{Orbit of the quantum walk}
Let the eigenvalues of the underlying twisted random walk be denoted by $\{\gamma_j(k)\}_{j=1}^{|V_0|}$. 
We define the orbit of the quantum walk by 
	\[ \Omega=\bigcup_{j=1}^{|V_0|} \Omega_j\subset \mathbb{R}^d, \]
where 
	\[ \Omega_j:=\{ \nabla\gamma_j(k) \;|\; k\in \mathcal{T}^d \}. \]
If we take the embedding of $L$ by $L=\{ n_1\hat{\theta}_1+\cdots+n_d\hat{\theta}_d \;|\; n_1,\dots,n_d \in \mathbb{Z} \}$, 
then the support of the continuous limit density function of the quantum walk $\rho_j$ is expressed by using $d\times d$ basis transformation matrix 
$\tilde{\Theta}:=[\hat{\theta}_1,\dots,\hat{\theta}_d]={}^t(\Theta^{-1})$ as 
	\[ \mathrm{supp}(\rho)=\tilde{\Theta}(\Omega). \]
\begin{theorem}\cite{HKSS}
Let $G$ be the $d$-dimensional lattice. Then we have 
	\[ \Omega \subseteq \{ x\in \mathbb{R}^d \;|\; ||x||^2 \leq 1/d \}. \]
\end{theorem}
In this paper, we newly obtain the orbits of the following crystal lattices. 
\begin{theorem}\label{main}
Let $G\in$ $\{$ triangular lattice, hexagonal lattice, kagome lattice $\}$ and 
$\Omega_G$ be the orbit of Grover walk on $G$. 
Then we have 
	\[ \Omega_G \subseteq \{ (x,y)\in \mathbb{R}^2 \;|\; x^2+{s(G)}xy+y^2\leq r(G) \}. \]
Here 
	\[ r(G)=\begin{cases} 
        		1/2 & \text{: $G$ is the triangular lattice,} \\ 
        		1/6 & \text{: $G$ is the hexagonal lattice,} \\
                        1/4 & \text{: $G$ is the kagome lattice,}
         \end{cases} \]
and 
	\[ s(G)=\begin{cases}
        	-1 & \text{: $G$ is the triangular lattice,} \\
              +1  & \text{: $G$ is the hexagonal lattice,} \\
              +1  & \text{: $G$ is the kagome lattice.}
        \end{cases} \]
\end{theorem}
\begin{corollary}
If we embed the above three lattices in $\mathbb{R}^2$ so that each euclidean length of edge is unit, then 
	\[ \Omega \subseteq \{ x\in \mathbb{R}^2 \;|\; ||x||^2\leq 1/2 \}.  \]
\end{corollary}
\begin{remark}
The opposite inclusion, that is, 
	\[ \Omega \supseteq \{ (x,y)\in \mathbb{R}^2 \;|\; x^2+{s(G)}xy+y^2\leq r(G) \} \] 
is an open problem except the $G=\mathbb{Z}^2$ case. 
We discuss it in the final section. 
\end{remark}
\begin{proof}
\noindent \\
\noindent{\bf The triangular lattice case.} \\
The twisted random walk on  the quotient graph of the triangular lattice is 
	\[ \hat{P}_k=\frac{1}{3}(\cos k_1+\cos k_2+\cos(k_1+k_2)).  \]
Thus its spectrum is 
	\[ \sigma(\hat{P}_k)=\frac{1}{3}(\cos k_1+\cos k_2+\cos(k_1+k_2))=\cos\gamma(k_1,k_2). \]
Then we have 
	\begin{equation}\label{xy1} 
        x=\frac{\partial\gamma}{\partial k_1}=\frac{\sin k_1+\sin (k_1+k_2)}{3\sin \gamma(k)},
        \; y=\frac{\partial\gamma}{\partial k_2}=\frac{\sin k_2+\sin (k_1+k_2)}{3\sin \gamma(k)}. 
        \end{equation}
We take the $-\pi/4$ rotation as  
	\[ u=\frac{x+y}{\sqrt{2}},\;v=\frac{-x+y}{\sqrt{2}}.  \]
Thus 
	\begin{equation}\label{uv1} 
        u=\frac{\sin k_1+\sin k_2+2 \sin(k_1+k_2)}{3\sqrt{2}\sin \gamma},\; v=\frac{-\sin k_1+\sin k_2}{3\sqrt{2} \sin \gamma}.  
        \end{equation}
Our target is to show 
	\[ \Omega'=\{ (u,v) \;|\; k_1,k_2\in \mathcal{T}\}=\{ (u,v)\in \mathbb{R}^2  \;|\; u^2+3v^2\leq 1\}. \]
We divide this proof into three steps as follows. \\
\begin{lemma}\label{step1}
 \[ \{(u,v)\in\mathbb{R}^2 \;|\; u^2+3v^2=1\}\subset \Omega'; \]
\end{lemma} 
\begin{lemma}\label{step2}
 \[ \{(u,v)\in\mathbb{R}^2 \;|\; u^2+3v^2\leq 1\}\supset \Omega'; \]
\end{lemma}
\begin{lemma}\label{step3}
 \[ \det(\mathrm{H}_\gamma) < \infty \]	
for almost every $k\in \mathcal{T}^2$.
\end{lemma}
\noindent Since the density of the limit distribution is expressed by the inverse of $\det(\mathrm{H}_\gamma)$, 
Lemma~\ref{step3} means the limit distribution takes positive values for almost every $(x,y)\in \Omega$. \\

\noindent{\bf Proof of Lemma~\ref{step1}}. 
When we take $(k_1,k_2)=0$, then both the numerator and the denominator of $u$ are $0$ and so are these for $v$. 
Then let us consider $k_1,k_2\ll 1$. 
Using the expansion of $\sin k_j\sim k_j$ and $\cos k_j\sim 1-k_j^2/2$ and taking $r=k_2/k_1$, (\ref{uv1}) is expressed by 
\begin{equation}
u\sim \pm \frac{3+3r}{\sqrt{12}\sqrt{1+r+r^2}}:=\tilde{u}_{\pm}(r), \; v\sim \pm \frac{-1+r}{\sqrt{12}\sqrt{1+r+r^2}}:=\tilde{v}_{\pm}(r). 
\end{equation}
It is easy to check that 
	\[ \tilde{u}_{\pm}(r)^2+3\tilde{v}_{\pm}(r)^2=1. \]
We have
	\[ \{ \tilde{u}_\pm(r) \;|\; r\in \mathbb{R}\}=[-1,1],\;\{ \tilde{v}_\pm(r) \;|\; r\in \mathbb{R}\}=[-1/\sqrt{3},1/\sqrt{3}]. \] 
Then the orbit of $(\tilde{u},\tilde{v})$ draws this ellipse, which completes the proof. $\square$ \\

\noindent{\bf Proof of Lemma~\ref{step2}}. 
We put $s_1=\sin k_1$, $s_2=\sin k_2$, $c_1=\cos k_1$, $c_2=\cos k_2$, $s_{12}=\sin(k_1+k_2)$, $c_{12}=\cos(k_1+k_2)$ and $c=\cos \gamma$, $s=\sin \gamma$. 
Now our target becomes to show 
	\begin{equation}\label{target1} 
        \left( \frac{s_1+s_2+2s_{12}}{3\sqrt{2}s} \right)^2+3\left( \frac{-s_1+s_2}{3\sqrt{2}s} \right)^2 \leq 1 
        \end{equation}
We repeat equivalent transformations of (\ref{target1}) as follows.  
	\begin{align*} 
        (\ref{target1})
                & \Leftrightarrow (s_1+s_2+2s_{12})^2+3(-s_1+s_2)-18+2(c_1+c_2+c_{12})^2 \leq 0 \\   
                & \Leftrightarrow -12+2s_1^2+2s_2^2+2s_{12}^2-4s_1s_2+4c_1c_2+4s_1s_{12}+4c_1c_{12}+4s_2s_{12}+4c_2c_{12}\leq 0 \\
                & \Leftrightarrow -12+2(s_1^2+s_2^2+s_{12}^2)+4(c_1c_2-s_1s_2)+4(c_1c_{12}+s_1s_{12})+4(c_2c_{12}+s_2s_{12})\leq 0 \\
                & \Leftrightarrow -6-2(c_1^2+c_2^2+c_{12}^2)+4(c_{12}+c_1+c_2)\leq 0 \\
                & \Leftrightarrow -2\{ (1-c_1)^2+(1-c_2)^2+(1-c_{12})^2 \} \leq 0 
        \end{align*}
Thus this completes the proof. \\

\noindent{\bf Proof of Lemma~\ref{step3}}. 
The determinant of $\mathrm{H}_\gamma$ is expressed by using $x$ and $y$ in (\ref{xy1})
	\[ \det(\mathrm{H}_\gamma)=\frac{(c_1+c_{12}-3cx^2)(c_2+c_{12}-3cy^2)-(c_{12}-3cxy)^2}{9s^2}. \]
It is obvious that the numerator of RHS is bounded. Thus only the case for $s=0$ has the possibility to provide $\det(\mathrm{H}_\gamma)=\infty$. 
\begin{align*} 
s=\sin\gamma=0 
	&\Leftrightarrow \cos\gamma= \pm 1 \\
        & \Leftrightarrow \frac{1}{3}(\cos k_1+\cos k_2+\cos(k_1+k_2))=\pm 1 \\
        & \Leftrightarrow k_1=k_2=0. 
\end{align*}
Thus the candidate of the place on $\mathbb{R}^d$ which produce $0$ as the value of the limit density function is on the ellipse. 
The Lebesgue measure of such a point is zero, which implies the conclusion.  \\

\noindent{\bf The hexagonal graph case.} \\
The twisted random walk on  the quotient graph of the hexagonal lattice is 
	\[ \hat{P}_k=\begin{bmatrix} 0 & \frac{1}{3}(\e^{-\im k_1}+\e^{-\im k_2}+1) \\ \frac{1}{3}(\e^{\im k_1}+\e^{\im k_2}+1) & 0  \end{bmatrix}. \]
Thus its spectrum is 
	\[ \sigma(\hat{P}_k)=\pm \frac{1}{3} |1+\e^{\im k_1}+\e^{\im k_2}|=\cos\gamma_{hex}(k_1,k_2). \]
We put $\gamma(k_1,k_2)=:\gamma_{hex}$
Then we have 
	\begin{equation}\label{xy} 
        x=\frac{\partial\gamma_{hex}}{\partial k_1}=\frac{2}{9}\frac{\sin k_1+\sin (k_1-k_2)}{\sin 2\gamma_{hex}},
        \; y=\frac{\partial\gamma_{hex}}{\partial k_2}=\frac{2}{9}\frac{\sin k_2+\sin (k_2-k_1)}{\sin 2\gamma_{hex}}. 
        \end{equation}
We take the $\pi/4$ rotation as  
	\[ u_{hex}=\frac{x-y}{\sqrt{2}},\;v_{hex}=\frac{x+y}{\sqrt{2}}.  \]
Thus 
	\begin{equation}\label{uv} 
        u_{hex}=\frac{\sqrt{2}}{9}\frac{\sin k_1-\sin k_2+2 \sin(k_1-k_2)}{\sin 2\gamma_{hex}},\; 
        v_{hex}=\frac{\sqrt{2}}{9}\frac{\sin k_1+\sin k_2}{\sin 2\gamma_{hex}}.  
        \end{equation}
Now our target becomes to show 
	\[ \Omega_{hex}'=\{ (u_{hex},v_{hex}) \;|\; (k_1,k_2)\in \mathcal{T}^2\}=\{ (u,v)\in \mathbb{R}^2  \;|\; 3u^2+9v^2\leq 1\}. \]
\begin{lemma}\label{step1H}
 \[ \{(u,v)\in\mathbb{R}^2 \;|\; 3u^2+9v^2=1\}\subset \Omega_{hex}'; \]
\end{lemma} 
\begin{lemma}\label{step2H}
 \[ \{(u,v)\in\mathbb{R}^2 \;|\; 3u^2+9v^2\leq 1\}\supset \Omega_{hex}'; \]
\end{lemma}
\begin{lemma}\label{step3H}
 \[ \det(\mathrm{H}_{\gamma_{hex}}) < \infty \]	
for almost every $k\in \mathcal{T}^2$.
\end{lemma}

\noindent{\bf Proof of Lemma~\ref{step1H}}. 
When we take $(k_1,k_2)=0$, then both the numerator and the denominator of $u$ are $0$ and also so are those of $v$. 
Then let us consider the case $k_1,k_2\ll 1$. 
Using the expansion of $\sin k_j\sim k_j$ and $\cos k_j\sim 1-k_j^2/2$ and taking $r=k_2/k_1$, (\ref{uv}) is expressed by 
\begin{equation}
u_{hex}\sim \pm\frac{1-r}{2\sqrt{1-r+r^2}}:=\tilde{u}_{\pm}(r), \; v_{hex}\sim \pm\frac{1+r}{6\sqrt{1-r+r^2}}:=\tilde{v}_{\pm}(r). 
\end{equation}
It is easy to check that 
	\begin{equation}\label{orbit00H} 
        \tilde{u}_{\pm}(r)^2+3\tilde{v}_{\pm}(r)^2=1/3. 
        \end{equation}
We have 
\[ \{ \tilde{u}_{\pm}(r) \;|\; r\in \mathbb{R}\}=[-1/\sqrt{3},1/\sqrt{3}],\;  \{ \tilde{v}_{\pm}(r) \;|\; r\in \mathbb{R}\}=[-1/3,1/3]. \] 
Then the orbit of $(\tilde{u},\tilde{v})$ draws this ellipse, which completes the proof. $\square$ \\

\noindent{\bf Proof of Lemma~\ref{step2H}}. 
We put $s_1=\sin k_1$, $s_2=\sin k_2$, $s_{12}=\sin(k_1-k_2)$, $c_{12}=\cos(k_1-k_2)$ and $c=\cos \gamma_{hex}$, $s=\sin \gamma_{hex}$. 
Our target is to show 
	\begin{equation}\label{target1H} 
        3\left( \frac{\sqrt{2}(s_1-s_2+2s_{12})}{18sc} \right)^2+9\left( \frac{\sqrt{2}(s_1+s_2)}{18sc} \right)^2 \leq 1 
        \end{equation}
We repeat equivalent transformations of (\ref{target1H}) as follows. 
	\begin{align*} 
        (\ref{target1H})  
        	& \Leftrightarrow (s_1-s_2+2s_{12})^2+3(s_1+s_2)^2\leq 54(sc)^2 \\
                & \Leftrightarrow 3(s_1-s_2+2s_{12})^2+9(s_1+s_2)^2\leq 4(3-(c_1+c_2+c_{12}))(3+2(c_1+c_2+c_{12})) \\
                & \Leftrightarrow s_1^2+s_2^2+s_{12}^2+(c_1c_2+c_2c_{12}+c_{12}c_1)-3\leq 0 \\
                & \Leftrightarrow -(c_1^2+c_2^2+c_{12}^2)+\frac{1}{2}((c_1+c_2+c_{12})^2-(c_1^2+c_2^2+c_{12}^2))\leq 0 \\
                & \Leftrightarrow (c_1+c_2+c_{12})^2-3(c_1^2+c_2^2+c_{12}^2)\leq 0. 
        \end{align*}
The Cauchy-Schwarz inequality implies the final inequality. 
Thus this completes the proof. \\

\noindent{\bf Proof of Lemma~\ref{step3H}}. 
Notice that $\partial \sin 2\gamma_{hex}/\partial k_j<\infty$ for $(j=1,2)$ by the above discussion. 
Therefore the determinant of $\mathrm{H}_{\gamma_{hex}}$ is expressed by using a bounded function $h(k_1,k_2)$ as 
	\[ \det(\mathrm{H}_{\gamma_{hex}})=\frac{h(k_1,k_2)}{\sin^4 2\gamma_{hex}}. \]
Only the case for $\sin 2\gamma_{hex}=0$ has a possibility to provide $\det(\mathrm{H}_\gamma)=\infty$. 
Such points on $\mathcal{T}^2$ are 
	\[ \{(2\pi/3,-2\pi/3),\;(-2\pi/3,2\pi/3),\;(0,0)\} \]
We have already examined the last case $(0,0)$; the corresponding orbit on $\mathbb{R}^2$ is the ellipse after the $\pi/4$ rotation (\ref{orbit00H}). 
The corresponding first and the second orbits with the $\pi/4$ rotation on $\mathcal{R}^2$ are obtained in the same way as the $(0,0)$ case: 
the orbits is computed by 
	\[ \{ (u,v) \;|\; u^2+3v^2= 1/6\}.  \]
Therefore the places on $\{ (u,v)\in \mathbb{R}^2 \;|\; u^2+3v^2\leq 1/3 \}$ where the value of the density may take zero is described by 
	\[ \{(u,v) \;|\; u^2+3v^2= 1/6 \;\mathrm{or}\; 1/3\}. \]
The above Lebesgue measure of the above set is zero. 
This completes the proof. \\

\noindent{\bf The kagome lattice case.} \\
The kagome lattice is the line graph of the hexagonal lattice. 
The transition matrix of the isotropic random walk on a graph $H$ is denoted by $P(H)$. 
We introduce the following well-known lemma. 
\begin{lemma}
Assume $H$ is a $\kappa$-regular graph. Let $L(H)$ be the line graph of $H$. Then we have 
	\[ \sigma(P(L(H)))=\varphi(\sigma(P(H))) \cup \{\frac{-1}{\kappa-1}\}, \] 
where $\varphi(x)=\frac{1}{2(\kappa-1)}(\kappa x+\kappa-2)$. 
Here the dimension of the eigenspace of $-1/(\kappa-1)$ is $|E(H)|-|V(H)|$. 
\end{lemma}
Thus the spectrum of the twisted random walk on the kagome lattice is described by 
	\begin{equation} \sigma_{0,kag}:=\{ \frac{1}{4}(1\pm |1+\e^{\im k_1}+\e^{\im k_2}|) \} \cup \{-\frac{1}{2}\} \end{equation}
We define $\cos \gamma_{kag}:=(1/4)+(3/4) \cos \gamma_{hex}\in \sigma_{0,kag}$. 
Then we have
	\begin{align*}
        \frac{\partial \gamma_{kag}}{\partial k_1} &= g(k_1,k_2) \frac{\partial \gamma_{hex}}{\partial k_1}, \\
        \frac{\partial \gamma_{kag}}{\partial k_2} &= g(k_1,k_2) \frac{\partial \gamma_{hex}}{\partial k_2},  
        \end{align*}
where 
	\[ g(k_1,k_2):=\frac{3}{4}\frac{\sin \gamma_{hex}}{\sin \gamma_{kag}}=\sqrt{3}\sqrt{\frac{1+\cos \gamma_{hex}}{5+3\cos \gamma_{hex}}}. \]
Taking the $\pi/4$ rotation, we have 
	\begin{equation}\label{uvkag} 
        u_{kag}=g(k_1,k_2)u_{hex},\;v_{kag}=g(k_1,k_2)v_{hex}. 
        \end{equation}
Remark that $0\leq g(k_1,k_2)\leq g(0,0)=\sqrt{3}/2$. Therefore using the previous fact on the hexagonal lattice, we have 
	\[ u_{kag}^2+3v_{kag}^2\leq 1/4. \]
The equality holds if and only if $(k_1,k_2)=(0,0)$. When $k_1,k_2\ll 1$ with $k_1/k_2=r$, then $g(k_1,k_2)=\sqrt{3}/2+O(||(k_1,k_2)||)$. 
Therefore we have 
	\[ \{ (u,v) \;|\; u^2+3v^2= 1/4 \}\subset \Omega'_{kag}\subset \{ (u,v) \;|\; u^2+3v^2\leq 1/4 \}  \]
The determinant of $\mathrm{H}_{\gamma_{kag}}$ is also expressed by using a bounded function $s(k_1,k_2)$ as 
	\[ \det(\mathrm{H}_{\gamma_{kag}})=\frac{s(k_1,k_2)}{(5+3\cos^2\gamma_{hex})\sin^4 2\gamma_{hex}}. \]
Only the case for $\sin 2\gamma_{hex}=0$ has a possibility of $\det(\mathrm{H}_{\gamma_{kag}})=\infty$. 
Thus we can use the argument for the previous hexagonal lattice case. 
This completes the proof. \\
\end{proof}
\noindent \\
\begin{figure}[htbp]
  \begin{center}
              \includegraphics[clip, width=7cm]{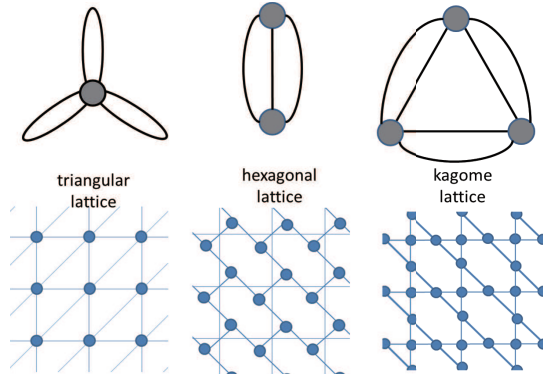}
              \caption{The quotient graphs and some realizations with $\hat{\theta}_1={}^t[1,0]$, $\hat{\theta}_2={}^t[0,1]$. }
  \end{center}
\end{figure}
\section{Summary and discussion}
We obtained the outer frames of the orbits of the Grover walker on some crystal lattices. 
We have shown that every orbit only depends on the embedding way of the fundamental lattices 
$L=\spann\{\hat{\theta}_1,\hat{\theta}_2\}$ in $\mathbb{R}^2$. 
If we choose the embedding way so that $\hat{\theta}_1={}^T[1,0]$, $\hat{\theta}_2={}^T[0,1]$, 
the outside frames are described by an ellipse. 
The natural question may arise about the interior. 
We have the following conjecture. 
\begin{conjecture}
Let $G\in$ $\{$ triangular lattice, hexagonal lattice, kagome lattice $\}$ and 
$\Omega_G$ be the orbit of Grover walk on $G$.
Then we have
\[ \Omega_G = \{ (x,y)\in \mathbb{R}^2 \;|\; x^2+{s(G)}xy+y^2\leq r(G) \}. \]
\end{conjecture}
The orbit $\Omega'_G$ after $s(G)\pi/4$ rotation is expressed by 
	\[ \Omega'_G=\{ (u(k_1,k_2),v(k_1,k_2)) \:|\: k_1,k_2\in[0,2\pi) \}, \]
where $u,v:[0,2\pi)^2\to \mathbb{R}$ are given by (\ref{uv1}) for ``$G=$ triangular lattice", (\ref{uv}) for ``$G=$ hexagonal lattice", 
and (\ref{uvkag}) for ``$G=$ kagome lattice". 
Figure~2 depicts the subregion of $\Omega'_G$ numerically which is the basis of our conjecture: 
 \[ \Omega_{r}=\{ (u(k,rk),v(k,rk)) \:|\: k\in[0,2\pi) \} \subset \Omega'_G \]
for $r=0,3,10,50,100$ cases. 
Note that $r$ corresponds to the pitch winding the torus $\mathcal{T}^2$, that is, 
the larger the pitch $r$ is, the more places of $\mathcal{T}^2$ ``$(k,rk)$" visits. 
\begin{figure}[htbp]
  \begin{center}
              \includegraphics[clip, width=15cm]{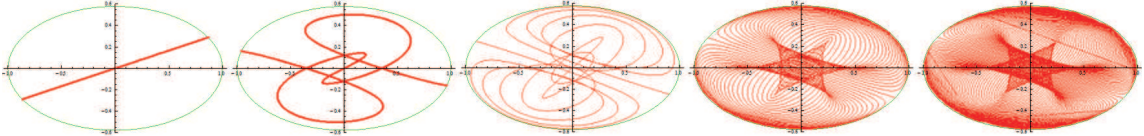}
              \caption{The orbits of $\Omega_r$ for the triangular lattice case by numerical simulation. ($r=0,3,10,50,100$)}
  \end{center}
\end{figure}
\begin{figure}[htbp]
  \begin{center}
              \includegraphics[clip, width=15cm]{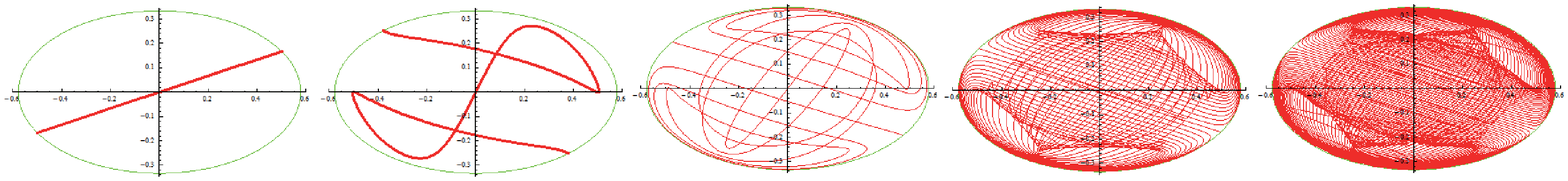}
              \caption{The orbits of $\Omega_r$ for the hexagonal lattice case by numerical simulation. ($r=0,3,10,50,100$) }
  \end{center}
\end{figure}
\begin{figure}[htbp]
  \begin{center}
              \includegraphics[clip, width=15cm]{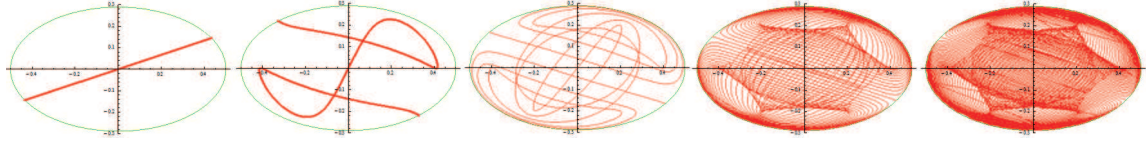}
              \caption{The orbits of $\Omega_r$ for the kagome lattice case by numerical simulation. ($r=0,3,10,50,100$) }
  \end{center}
\end{figure}
%

%


\par
\
\par\noindent
\noindent
{\bf Acknowledgments.}
\par

N.K. is partially supported by the Grant-in-Aid for Scientific Research (Challenging Exploratory Research) 
of Japan Society for the Promotion of Science (Grant No. 15K13443).
E.S. acknowledges financial supports from the Grant-in-Aid for Young Scientists (B) and of Scientific Research (B) Japan
Society for the Promotion of Science (Grant No. 16K17637, No. 16K03939). 
The research by H.J.Y. was supported by Basic Science Research Program through the National
Research Foundation of Korea (NRF) funded by the Ministry of Education (NRF-2016R1D1A1B03936006).
\par

\begin{small}
\bibliographystyle{jplain}

\end{small}



\end{document}